\documentclass[preprint,amsmath,amssymb,aps]{revtex4-1}

\usepackage{graphicx}
\usepackage{dcolumn}
\usepackage{bm}

\begin{document}


\title{Finsler pp-waves}

\author{Andrea Fuster}
 \altaffiliation{A.Fuster@tue.nl}

\affiliation{
 Eindhoven University of Technology \\
 Eindhoven, The Netherlands
}

\author{Cornelia Pabst}
\altaffiliation{pabst@strw.leidenuniv.nl}

\affiliation{
Universiteit Leiden \\
 Leiden, The Netherlands
}
%
%


\date{\today}

\begin{abstract}
In this work we present a Finslerian version of the well-known pp-waves, which generalizes the very special relativity (VSR) line element. Our Finsler pp-waves are an exact solution of Finslerian Einstein's equations in vacuum. 
\end{abstract}

\maketitle






\newpage
\section{Introduction}
Finsler geometry is a generalization of Riemannian geometry in which geometrical quantities are direction dependent. The main object is the so-called Finsler structure $F$, which defines the infinitesimal line element $ds=F$. Riemannian geometry is recovered when the (square of the) Finsler function is constrained to be quadratic, $F(x,dx)=\sqrt{g_{ij}(x)dx^idx^j}$. Finsler geometry has found applications in several fields of research where anisotropic media play a role, such as seismology \cite{antonelli}, optics \cite{duval} and medical imaging \cite{melonakos,florack}.   Finsler geometry is also appealing in relativity and cosmology, in particular concerning scenarios violating full Lorentz invariance.  \\

In this work we consider the very special relativity (VSR) framework proposed by Cohen and Glashow \cite{Cohen}, where only a subgroup of the full Lorentz group is preserved. Soon thereafter, Gibbons et al. \cite{Gibbons} pointed out the Finslerian character of the corresponding line element, an anisotropic generalization of Minkowski spacetime. The question arises of whether it is possible to find anisotropic generalizations of curved spacetimes, which satisfy Finslerian Einstein's equations. Several approaches to Finslerian extensions of the general relativity equations of motion have been suggested \cite{Voicu,Li,Pfeifer}. In this work we choose to adopt the framework by Pfeifer and Wohlfarth \cite{Pfeiferjournal,Pfeifer}, in which the Finslerian field equation is derived from a well-defined action and the geometry-related term obeys the same conservation law as the matter source term. \\

Finslerian extensions of well-known exact solutions such as the Schwarzschild metric have been proposed \cite{silagadze}. In the context of cosmology, Finslerian versions of the Friedman-Robertson-Walker metric have been recently studied \cite{kouretsis,stavrinos}. We propose a Finslerian version of the well-known pp-waves. The Riemannian pp-waves belong to a wider class of spacetimes with the property that all curvature invariants of all orders vanish, the so-called vanishing scalar invariant (VSI) spacetimes \cite{coley}. These are relevant in some supergravity and string scenarios since they are, due to the VSI property, exact solutions to the corresponding equations of motion \cite{higherVSI,higherVSIsugra,coley-gibbons}. We show that our Finsler pp-waves are an exact solution of Finslerian Einstein's equations in vacuum.

\newpage
\section{Theory}
\subsection{(Pseudo-)Finsler geometry in a nutshell}

Let $M$ be an $n$-dimensional $C^{\infty}$ manifold. We denote the tangent bundle of $M$, the set of tangent spaces $T_xM$ at each $x\in M$, by $TM := \{T_xM\vert x\in M\}$. We can write each element of $TM$ as $(x,y)$, where $x\in M$ and $y\in T_xM$.  \\

A Finsler structure is a function defined on the tangent bundle $TM$ 
\begin{align}
F: TM \rightarrow [0,\infty)
\end{align}
satisfying the following properties:
\begin{enumerate}
\item \emph{Regularity}: $F$ is $C^{\infty}$ on the slit tangent bundle $TM\setminus 0 = TM\setminus\{y=0\}$. \item \emph{Homogeneity}: $F(x,\lambda y)=\lambda F(x,y), \quad \forall\, \lambda>0$ and $(x,y)\in TM$. 
\item \emph{Strong convexity}: The fundamental metric tensor 
\begin{equation}\label{fundamentaltensor}
g_{ij}(x,y)=\frac{1}{2} \frac{\partial^2 F^2 (x,y)} {\partial y^i \partial y^j} 
\end{equation}
with $i,j=1,\dots,n$, is positive definite for all $(x,y)\in TM\setminus 0$.
\end{enumerate}
The pair $(F,M)$ is called a Finsler manifold or Finsler space. A Finsler manifold is Riemannian when the fundamental tensor is independent of the tangent vector $y$, $g_{ij}(x) \equiv g_{ij}(x,y)$.  
A thorough treatment of Finsler geometry can be found in \cite{Baoetal}. To be precise, in this work we regard pseudo-Finsler spaces, for which the regularity property may not hold on the zero vectors and the fundamental tensor is \emph{not} restricted to be positive definite.  

\subsection{Finslerian generalization of Einstein's equations}\label{FinslerGR}
In this section we treat one of the approaches to construct a Finslerian version of Einstein's equations. In doing so we consider the particular case of vacuum and Berwald spaces, which we characterize later. \\

The geodesic spray coefficients are defined as
\begin{equation} \label{geodspray}
G^i := \gamma^i\,_{jk} y^j y^k 
\end{equation}
where $\gamma^i\,_{jk}$ are the Christoffel symbols of the second kind. These adopt the same form as in the Riemannian case, but in this setting we generally have $\gamma^i\,_{jk}=\gamma^i\,_{jk}\, (x,y)$. As their name suggests the geodesic spray coefficients play a role in the Finslerian geodesic equations, which in the simplest case take the form 
\begin{equation}
\ddot{x}^i+G^i=0
\end{equation}
with $\ddot{x}^i=\partial^2 x^i/\partial t^2$, and $t$ parameterizing the geodesic curve. From these, the tensor known as the predecessor of the flag curvature can be constructed as follows
\begin{align}
R^i\,_k = \frac{1}{F^2}\left(\partial_{x^k}G^i - \frac{1}{4}\partial_{y^k}G^j\partial_{y^j}G^i - \frac{1}{2}y^j\partial_{x^j}\partial_{y^k}G^i + \frac{1}{2}G^j\partial_{y^j}\partial_{y^k}G^i\right)
\end{align}
where $\partial_{x^k}=\partial / \partial x^k$, $\partial_{y^k}=\partial / \partial y^k$. This curvature tensor certainly relates to the Finslerian Riemann tensor, but it does \emph{not} reduce to the Ricci tensor in the Riemannian limit. \\

We are now ready to introduce the simplest curvature scalar in Finsler geometry, the Finslerian Ricci scalar, as the trace of tensor $R^i\,_k$ \cite{Baoetal}:
\begin{align}
Ric := R^i\,_i
\end{align}
Clearly, its construction is notably different from that of the Ricci scalar in Riemannian geometry. On the other hand, the following Finslerian Ricci tensor has been proposed \cite{akbar}:
\begin{align}
Ric_{ij} := \frac{1}{2}\partial_{y^i}\partial_{y^j}(F^2 Ric)
\end{align}
This tensor is manifestly covariant and symmetric. Moreover, if the Finsler structure $F$ is Riemannian, it reduces to the usual Ricci tensor. It is therefore a natural generalization of the Ricci tensor. 
In what follows we adopt the approach by Pfeifer and Wohlfarth \cite{Pfeiferjournal,Pfeifer}, where a Finslerian Einstein-Hilbert action is proposed and a vacuum Finslerian scalar field equation is derived (not reproduced here). \\

\textbf{Berwald spaces.}
We consider now a particular class of Finsler spaces, the  so-called \emph{Berwald spaces}. These are defined by the property that the geodesic spray coefficients $G^i$, Eq. (\ref{geodspray}), are quadratic in $y$. In this case we obtain the Finslerian vacuum equation \cite{prep} 
\begin{equation}
f^{kl}_{ij}(g,l,A)\;Ric_{kl}=0  \label{Finslervacuumeq}
\end{equation}
where $f^{kl}_{ij}$ is a (2,2)-tensor constructed from the fundamental tensor, the distinguished section $l^i=y^i/F$ and the Cartan tensor $A_{ijk}=(F/2)\;\partial_{y^k} g_{ij}$ and its covariant derivatives. We derive this equation by restricting Pfeifer's Finslerian field equation to the Berwald case. \\

Therefore, any Berwald metric satisfying 
\begin{equation}
Ric_{ij} = 0   \label{FinslerEE}
\end{equation}
is an exact solution of the Finslerian vacuum equation (\ref{Finslervacuumeq}). The condition above is formally the same as in the Riemannian case.   

\subsection{Very special relativity}
Cohen and Glashow \cite{Cohen} pointed out that the local laws of physics do not need to be invariant under the full Lorentz group but only under a certain subgroup, ISIM(2). 
This is called Very Special Relativity (VSR). Subsequently, Gibbons, Gomis and Pope \cite{Gibbons} studied  deformations of the subgroup ISIM(2), investigating possibilities to incorporate gravity into the theory. They showed that the 1-parameter family $\mbox{DISIM}_b(2)$, where D stands for deformation and $b$ is a dimensionless parameter, is the only physically acceptable deformation of the VSR subgroup. The line element which is invariant under $\mbox{DISIM}_b(2)$ reads  
\begin{equation}\label{VSR}
ds=\left(-2dudv+(dx^1)^2+(dx^2)^2\right)^{(1-b)/2}(-du)^b
\end{equation}
where we use coordinates $x^{\mu}=(u,v,x^1,x^2)$, and $u=(1/\sqrt{2})(t-x^3)$, $v=(1/\sqrt{2})(t+x^3)$ are light-cone coordinates. Obviously, this reduces to the Minkowski line element for $b=0$. As Gibbons et. al. noted, (\ref{VSR}) is a Finslerian line element, which had already been suggested by Bogoslovsky \cite{Bogos}. The associated Finsler structure is given by
\begin{equation}
F(x,y) = \left(-2y^uy^v  + (y^1)^2 + (y^2)^2\right)^{(1-b)/2}(-y^u)^{b}
\end{equation}
where $y^{\mu}=(y^u,y^v,y^1,y^2)$. 
This space has vanishing geodesic spray coefficients  
\begin{equation}
G^i=0
\end{equation}
and therefore Finslerian curvature tensors are zero, and  
geodesics are defined by linear equations. Such a space is called projectively flat.   
\subsection{Finsler pp-waves}
Let us now consider the spacetime: 
\begin{equation}\label{curvedVSR}
ds=(g_{\mu\nu}dx^{\mu}dx^{\nu})^{(1-b)/2}(-du)^b
\end{equation}
This is a modification of the Bogoslovsky line element given by Eq. (\ref{VSR}), where the Minkowski metric is substituted by a general Riemannian metric $g_{\mu\nu}$. We propose the case where $g_{\mu\nu}\equiv \mbox{pp-waves}$. The well-known pp-wave metric is given by \cite{pp-waves}:
\begin{equation}
ds^2=-2dudv - \Phi(u,x^1,x^2) du^2 + (dx^1)^2 + (dx^2)^2
\end{equation}
This metric is an exact vacuum solution of the Einstein's equations when the function $\Phi$ is harmonic, $\triangle_{(x^1,x^2)}\Phi=0$, and it describes exact gravitational waves propagating in a Minkowski background. Thus we have the following ansatz:
\begin{equation}\label{VSRpp}
ds=\left(-2dudv- \Phi(u,x^1,x^2) du^2 + (dx^1)^2+(dx^2)^2\right)^{(1-b)/2}(-du)^b
\end{equation}
Note that we recover the pp-wave metric in the case $b=0$. The associated Finsler structure is of the form:
\begin{equation} \label{FinslerppF}
F(x,y) = \left(-2y^uy^v - \Phi(u,x^1,x^2) (y^u)^2 + (y^1)^2 + (y^2)^2\right)^{(1-b)/2}(-y^u)^{b}
\end{equation}
From $F(x,y)$ we derive the fundamental tensor (see Appendix), the Christoffel symbols and the corresponding geodesic spray coefficients 
defined by Eq. (\ref{geodspray}): 
\begin{align}
G^u &= 0 \nonumber \\
G^v &= y^uy^1\partial_{x^1}\Phi + y^uy^2\partial_{x^2}\Phi + \frac{(y^u)^2}{2}\partial_{u}\Phi  \nonumber \\
G^1 &= \frac{(y^u)^2}{2} \partial_{x^1}\Phi  \nonumber \\
G^2 &= \frac{(y^u)^2}{2} \partial_{x^2}\Phi \nonumber
\end{align}
These are obviously quadratic in $y$, and thus the postulated Finslerian space is of Berwald type. Recall from section \ref{FinslerGR} that the vanishing of the Finslerian Ricci tensor is a sufficient condition for a Berwald space to be an exact Finsler vacuum solution. \\

The Finslerian Ricci scalar can be computed to be:
\begin{align}
Ric = \frac{(y^u)^2}{2F^2} \Big(\partial_{x^1}^2\Phi+\partial_{x^2}^2\Phi\Big)
\end{align}
Hence, the Finslerian Ricci tensor has the only non-zero component:
\begin{align}
Ric_{uu} = \frac{1}{2}(\partial_{x^1}^2\Phi+\partial_{x^2}^2\Phi)=\frac{1}{2}\triangle_{(x^1,x^2)}\Phi
\end{align}
Therefore, we conclude that every harmonic function $\Phi$ leads to an exact solution of the Finslerian Einstein equations in vacuum, Eq. (\ref{FinslerEE}). This is completely analogous to the Riemannian case. Interestingly, an harmonic function $\Phi$ also leads to a solution of the vacuum Finslerian Einstein's equations proposed in \cite{Li}, which are of the form $Ric=0$.  \\

As a curiosity, we point out the Finslerian analogue of the scalar curvature 
\begin{align}
R=g^{uu}Ric_{uu}=\frac{2b}{1+b}Ric
\end{align}
which vanishes as expected in the Riemannian limit $b=0$. \\

\textbf{Remark}. We point out that the fundamental tensor arising from line element (\ref{curvedVSR}) is ill-defined in the subspaces $y^u=0$ and $g_{\mu\nu}y^{\mu}y^{\nu}=0$, the null cone of the Riemannian metric. In such cases the Finsler structure (\ref{FinslerppF}) vanishes and the fundamental tensor becomes singular, $\mbox{det}(g_{ij})=0$. However, this is not specific to our Finsler pp-waves since it holds for the Bogoslovsky line element (\ref{VSR}) as well, in the subspaces $y^u=0$ and $\eta_{\mu\nu}y^{\mu}y^{\nu}=0$, with $\eta_{\mu\nu}$ the Minkowski metric. This shortcoming of pseudo-Finslerian line elements of this type has already been discussed in \cite{Pfeifer}.
\section{Discussion}
In this work we present a Finslerian version of the well-known pp-waves, which generalizes the (deformed) VSR line element investigated by Gibbons et. al. Our Finsler pp-waves are an exact solution of Finslerian Einstein's equations in vacuum. \\

In future work we will consider a number of open issues. First of all, the physical implications of the singularities of the fundamental tensor should be investigated. It would also be interesting to study the curvature scalar invariants corresponding to the presented Finsler pp-waves solution.
Recall that Riemannian pp-waves belong to the class of spaces with vanishing scalar invariants. However, it is not clear how such spaces would be defined in a Finsler setting since scalar quantities, such as the Finslerian Ricci scalar and the Cartan scalar, are in principle direction-dependent.   
Last, an obvious next step would be to consider solutions of Finslerian VSI type, where generalizations of the pp-wave metric are employed, both in four and in higher dimensions.   
\section*{Acknowledgments}
A. Fuster would like to acknowledge C. Pfeifer for email correspondence, T. Dela Haije for proofreading the manuscript and A. Ach\'{u}carro for everything she did to make this paper happen. 

\section*{Appendix}
The components of the fundamental tensor $g_{ij}$ corresponding to ansatz (\ref{FinslerppF}) read as follows, with $\alpha=\left(-2y^uy^v-\Phi(u,x^1,x^2)\;(y^u)^2+(y^1)^2+(y^2)^2\right)$:
\begin{align}
g_{uu} 
&= \frac{F^2}{\alpha}\left(-\Phi - b \left(\frac{((y^1)^2+(y^2)^2)^2-2b ((y^1)^2+(y^2)^2-y^uy^v)^2}{\alpha (y^u)^2} - \frac{2 (y^v)^2 - (y^u)^2\Phi}{\alpha} + \Phi\right)\right) \nonumber \\
g_{uv} 
&= \frac{F^2}{\alpha}\left(-1 -b\left((1-2b)+2(1-b)\frac{y^uy^v+(y^u)^2\Phi}{\alpha}\right)\right)\nonumber \\
g_{vv} &= \frac{F^2}{\alpha}(-2b)(1-b)\frac{(y^u)^2}{\alpha} \nonumber \\
g_{11} &= \frac{F^2}{\alpha}(1 - b) \left(1-2 b \frac{(y^1)^2}{\alpha}\right) \nonumber \\
g_{22} &= \frac{F^2}{\alpha}(1 - b) \left(1-2 b \frac{(y^2)^2}{\alpha}\right) \nonumber \\
g_{u1} &= \frac{F^2}{\alpha}2b (1 - b)\left(\frac{y^1}{y^u}+ \frac{ y^1(y^v+y^u\Phi)}{\alpha}\right) \nonumber \\
g_{u2} &= \frac{F^2}{\alpha}2b (1 - b)\left(\frac{y^2}{y^u}+ \frac{ y^2(y^v+y^u\Phi)}{\alpha}\right) \nonumber \\
g_{v1} &= \frac{F^2}{\alpha}2b(1-b) \frac{y^1y^u}{\alpha} \nonumber \\
g_{v2} &= \frac{F^2}{\alpha}2b(1-b) \frac{y^2y^u}{\alpha} \nonumber
\end{align}

\end{document}